\documentclass[english]{article}
\usepackage[LGR,T1]{fontenc}
\usepackage{xcolor}
\usepackage{array}
\usepackage{amstext}
\usepackage{stackrel}
\usepackage{graphicx}
\usepackage{esint}

\makeatletter

\DeclareRobustCommand{\greektext}{%
  \fontencoding{LGR}\selectfont\def\encodingdefault{LGR}}
\DeclareRobustCommand{\textgreek}[1]{\leavevmode{\greektext #1}}
\ProvideTextCommand{\~}{LGR}[1]{\char126#1}
\providecommand{\keywords}[1]
{
  \small	
  \textbf{\textit{Keywords---}} #1
}
\providecommand{\tabularnewline}{\\}
\usepackage[a4paper]{geometry}
\makeatother

\usepackage{babel}
\begin{document}
\title{Higher-Order Nonclassicality in Photon Added and Subtracted Qudit
States}
\author{Kathakali Mandal and Amit Verma\\
Jaypee Institute of Information Technology, Sector-128, Noida, UP-201304,
India}
\maketitle
\begin{abstract}
Higher-order nonclassical properties of $r$ photon added 
and $t$ photon subtracted qudit states (referred to as $rPAQS$ and $tPSQS$, respectively) are investigated here to answer: How addition and subtraction of photon can be used engineer higher-order nonclassical properties of qudit states? To obtain the  answer higher-order moment of relevant bosonic field operators are first obtained for  $rPAQS$ and $tPSQS$, and subsequently the same is used to study the higher-order nonclassical properties of the corresponding states. A few witnesses of higher-order nonclassicality (e.g. Higher-Order Antibunching, Higher-Order Squeezing of
Hillery  type, Higher-Order sub-Poissonian Photon
statistics) are first used to establish that $rPAQS$ and $tPSQS$ are highly nonclassical. These witnesses are found to indicate that the amount of nonclassicality enhances with the number of photon added ($r$). To quantitatively establish this observation and to make a comparison between $rPAQS$ and $tPSQS$,  volumes of the negative part of Wigner function (nonclassical volume) of $rPAQS$ and $tPSQS$ are computed as a quantitative measure of nonclassicality. Finally, for the sake of verifiablity of the obtained results, optical tomograms are also reported which can be obtained experimentally and used to produce Wigner function (which is not directly measurable in general) by the Radon transform.  Throughout the study, a particular type of qudit state named as a new generalized binomial state is used as an example. 
\end{abstract}

\section{Introduction}

Various exciting applications of nonclassical properties of quantum
states have recently been proposed and realized. Specifically, squeezed states have been used
in the LIGO experiment for the detection of gravitational waves \cite{abbott2016ligo,abbott2016gw151226} and in
continuous variable quantum key distribution \cite{gottesman2003secure,cerf2001quantum,madsen2012continuous,weedbrook2012gaussian};  entangled states have been used in quantum teleportation and quantum cryptography
\cite{bennett1993teleporting,ekert1991quantum,bennett1992quantum}
moreover antibunching is found to be useful in characterizing single photon sources
\cite{pathak2010recent,verma2008higher}. In quantum state engineering
\cite{marchiolli2004engineering,miranowicz2004dissipation,vogel1993quantum,sperling2014quantum}
and quantum computation \cite{nielsen2002quantum,pathak2013elements},
nonclassical properties of any quantum state are getting much attention \cite{barnett2018statistics,verma2010generalized,pathak2014wigner,verma2008higher,verma2019study,alam2019bose}.
This is so because nonclassical states have no classical analogue
and can thus be useful in realizing tasks that are impossible
in the classical world. In other words, nonclassical states which
are characterized by the negative values of Glauber-Sudarshan $P$-function,
can only establish quantum supremacy \cite{harrow2017quantum,neill2018blueprint}.
Examples of nonclassical properties are squeezing, antibunching and
entanglement. In summary, nonclassical features of quantum states
are very important and the same has been studied for various families
of quantum states \cite{verma2019study,mandal2019generalized}. One such family of quantum states is called Qudit ($d$-level states)
states \cite{verma2010generalized,pathak2014wigner,verma2008higher} which may be viewed as finite superposition of Fock states in $d$-dimensional
Hilbert space. A particular subclass of qudit states is the set of finite dimensional intermediate states. These states are interesting because any state of this
family can be reduced to various other quantum states at different
limits of the state parameters. Usually, intermediate states correspond to states having photon number distribution analogous to a well defined statistical distribution
function for like Binomial function, 
negative binomial function and many more. The intermediate state is named in accordance with statistical distribution that represents the photon number distribution of the state.  One such intermediate state  named as new generalized binomial state $(NGBS)$ was
introduced by Fan et al \cite{fan1999new} and we have recently reported
 higher-order nonclassicality in NGBS \cite{mandal2019generalized}. Here we aim to extend the work and check how addition and subtraction of photon can engineer the higher-order nonclassical properties of this state. In short, in what follows, we aim to study higher-order nonclassical properties of photon added and subtracted NGBS.

 The nonclassical properties we aim to study, can be witnessed by some well defined
inequality in terms of moments of creation and annihilation operators
under the framework of second quantization. These criteria can have
lower-order and higher-order versions. Study of lower-order nonclassicality
of a quantum state is reported in literature since the early days of quantum optics, but
interest in higher-order nonclassicality is relatively new and promising for the experiments
point of view also as it can detect weaker nonclassicality which are hard to observe using lower-order criteria \cite{allevi2012high,allevi2012measuring}.
\begin{figure}
\begin{center}
\includegraphics[scale=0.7]{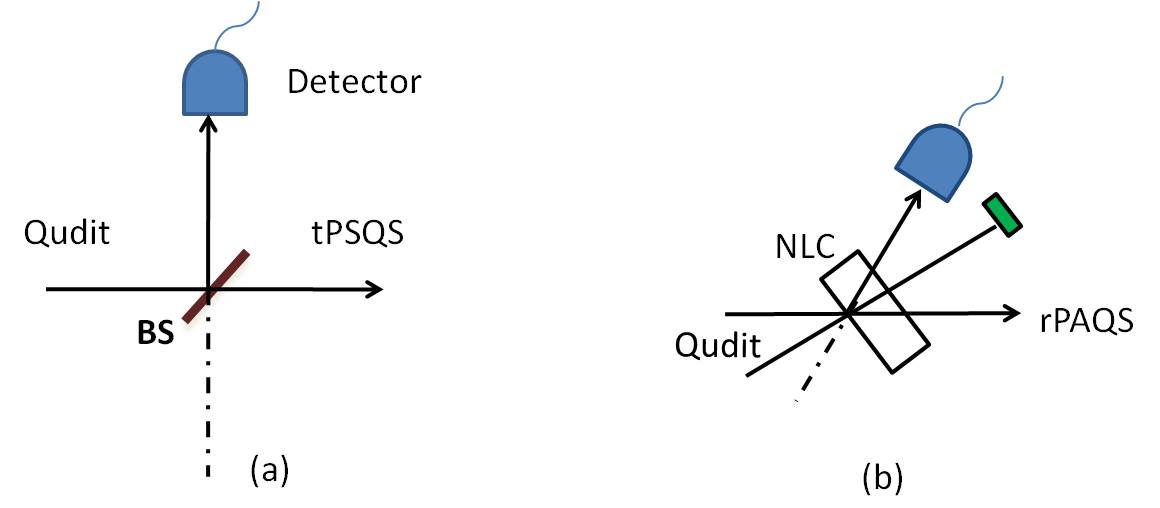}
\caption{(color online) A schematic diagram for generating photon subtracted and photon added qudit state in (a) $tPSQS$ by using a beam splitter and
(b) $rPAQS$ by using nonlinear crystal, respectively.}
\end{center}
\end{figure}

We have already mentioned that in what follows we wish to study higher-order nonclassical properties of photon added and subtracted NGBS.
Now the question arises, how can we generate these (photon added and/or subtracted) nonclassical states? Here, to address this question, we wish to note that any nonclassical state
can be generated by two kinds of operations, unitary and nonunitary
operations. In unitary operation, nonclassical state is generated
under control of a Hamiltonian and the example of nonunitary operation
is like photon addition and subtraction to a quantum state. Here in
\(Fig. 1\), we have provided the possible scheme for experimental realization
of photon addition and subtraction to a quantum state. Operation of
photon subtraction and addition is depicted by using a beam splitter
\(Fig. 1(a)\) and a nonlinear crystal \(Fig. 1(b)\), respectively. These strategies can be used to generate engineered quantum states. Interesting
examples of such engineered nonclassical states are Fock state, photon
added/subtracted coherent state, displaced Fock state, intermediate
state like binomial state (BS) and New generalized binomial state (NGBS). In our earlier works, we have reported higher-order nonclassicality in different quantum systems
\cite{mandal2019generalized,verma2008higher,verma2010generalized, verma2019study}. However, the effect of multiple photon addition and subtraction on qudits in general and NGBS in particular have not yet been studied. Although it can reveal higher-order nonclassical properties of a family of quantum states in different limits as photon added and subtracted NGBS can be reduced to various states in different limits. We will return to this point later. Here we just wish to note that this particular possibility, the above mentioned features and  aspects of higher-order nonclassicality have motivated us for the present study and in what follows we will study higher-order nonclassical properties of $r$ photon added qudit state ($rPAQS$) and $t$ photon subtracted qudit state ($tPSQS$).

Higher-order nonclassicality can be witnessed through various operational criteria (inequalities), most of them are expressed as a function of moments of annihilation and creation operators. Keeping that in mind, in what follows, we have  first expressed $rPAQS$ and  $tPSQS$ in general as Fock superposition states, and subsequently used that to obtain expressions for a general moment of annihilation and creation operator (say, \(\langle a^{\dagger k}a^{l}\rangle \))  which in turn provides us analytic expression for the nonclassicality witnessing parameters for various higher-order nonclassical phenomena, e.g.,  Higher-Order Antibunching (HOA), Higher-Order Squeezing (HOS-Hillery type), Higher-Order sub-Poissonian Photon statistics (HOSPS), etc. The systematic study revealed that the depth of nonclassicality witnessing parameter increases with the number of added photon, but no conclusive decision can be made from them as none of the witnesses of nonclassicality can yield a quantitative measure. So we looked back to a quasi-distribution function, namely Wigner function whose negative parts illustrate the presence of nonclassicality and volume of the negative part quantifies nonclassicality. The quantification helped us to establish that nonclassicality indeed increases with the addition of photon, it further helped us compare  $rPAQS$ and  $tPSQS$. This was consistent with the witness based observation, but still an issue remained- Wigner function is not measurable in general. So to complete the work, we have computed optical tomograms for the quantum states of our interest. Optical tomograms can be produced experimentally, and thus, they can be used to verify our results and subsequently Radon transform can be used to obtain Wigner function from optical tomogram. This part makes the predictions of present analytic study experimentally verifiable. Before we proceed to the more technical part of the paper, it will be apt to note that in the first part of the paper a general construction of the problem is done in terms of $rPAQS$ and  $tPSQS$, but the higher-order nonclassicality witness are illustrated by considering a particular type of qudit state only (namely NGBS state \cite{fan1999new}).Thus, in what follows, higher-order nonclassicality will be witnessed and quantitatively measured for $r$ photon added NGBS and $t$ photon subtracted NGBS.  Earlier we reported higher-order nonclassicality in NGBS \cite{mandal2019generalized}, clearly those results will be obtained as special cases of the present results with $r=0$ and $t=0$. Further, in an earlier study  nonclassical properties of single photon added and subtracted in binomial state \cite{mandal2019higher} were studied. One can easily understand that the present results would be so general that all such existing results will be reducible from the present results at different limits. In addition to our earlier works, a large number of works have recently been performed to elaborate on the relevance and importance of single photon and multi photon addition and subtraction in different quantum states (see \cite{malpani2020impact, li2019multiple, zhang2020nonclassicality, wang2019time,ren2019nonclassicality} and references therein). Most these works were focused to specific quantum states, and that's set the motivation of the present work where aim to approach the problem from a much more general perspective as far as the state to be considered and the expression for moment of field operators are concerned.

This paper is organized in 5 sections. In section 2, we present the
mechanism of photon addition and subtraction in general qudit state
and further, we provide expressions of higher-order moment for studying
various higher-order nonclassical phenomena. In section 3, various
criteria of witness of higher-order nonclassicality are explored. Section
4 provides quantitative analysis of higher-order nonclassicality in
the form of Nonclassical volume.
Finally section 5 conclude our results.

\section{Photon addition and subtraction in qudit state}
A qudit state of radiation field in Fock basis can be expressed as 
\begin{equation}
|\psi\rangle=\stackrel[n=0]{M}{\sum}C_{n}|n\rangle,\label{eq:psi}
\end{equation}
 where $C_{n}$ is the probability amplitude and $|n\rangle$ represents
a Fock state having $n$ photon. 
If we add $r$ photon to this state through creation operator, we
obtain a new qudit state as $r$ photon
added qudit state and can be expressed as
\begin{equation}
a^{\dagger r}|\psi\rangle=|rPAQS\rangle=N_{r}\stackrel[n=0]{M}{\sum}C_{n}\left[\frac{\left(n+r\right)!}{n!}\right]^{1/2}|n+r\rangle,\label{eq:rPABS}
\end{equation}
 where $N_{r}$ is normalization constant and is defined as
\begin{equation}
N_{r}=\left[\stackrel[n=0]{M}{\sum}C_{n}^{2}\frac{\left(n+r\right)!}{n!}\right]^{-1/2}.\label{eq:nr}
\end{equation}
Similarly a $t$ photon subtracted qudit state can be obtained by repeatedly applying annihilation operator and can be expressed as
\begin{equation}
a^{t}|\psi\rangle=|tPSQS\rangle=N_{t}\stackrel[n=0]{M}{\sum}C_{n}\left[\frac{n!}{\left(n-t\right)!}\right]^{1/2}|n-t\rangle,\label{eq:tPSBS}
\end{equation}
where $N_{t}$ is normalization constant and defined as

\begin{equation}
N_{t}=\left[\stackrel[n=0]{M}{\sum}C_{n}^{2}\frac{n!}{\left(n-t\right)!}\right]^{-1/2}.\label{eq:Nt}
\end{equation}
Now to study nonclassical properties of
theses states, we would required analytic expressions of the moments
of the relevant field operators. For $r$ photon added qudit state, a bit
of computation would yield

\begin{equation}
\langle a^{\dagger k}a^{l}\rangle_{rPAQS}=\left|N_{r}\right|^{2}\stackrel[n=|l-k|]{M}{\sum}\left[C_{n}^{2}C_{n-l+k}^{2}\frac{\left(n+s\right)!^{2}\left(n+s-l+k\right)!^{2}}{n!\left(n-l+k\right)!\left(n+s-l\right)!^{2}}\right]^{1/2},l>k\label{eq:mb}
\end{equation}
and
\begin{equation}
\left\langle a^{\dagger k}a^{l}\right\rangle _{rPAQS}=\left|N_{r}\right|^{2}\stackrel[n=0]{M-|l-k|}{\sum}\left[C_{n}^{2}C_{n-l+k}^{2}\frac{\left(n+s\right)!^{2}\left(n+s-l+k\right)!^{2}}{n!\left(n-l+k\right)!\left(n+s-l\right)!^{2}}\right]^{1/2},l\leq k.\label{eq:mc}
\end{equation}
Similarly, we can get analytic expressions of the moments  for $t$ photon subtracted qudit state

\begin{equation}
\langle a^{\dagger k}a^{l}\rangle_{tPSQS}=\left|N_{t}\right|^{2}\stackrel[n=|l-k|]{M}{\sum}\left[C_{n}^{2}C_{n-l+k}^{2}\frac{n!\left(n-l+k\right)!}{\left(n-t-l\right)!}\right]^{1/2},l>k\label{eq:me}
\end{equation}
and
\begin{equation}
\left\langle a^{\dagger k}a^{l}\right\rangle _{tPSQS}=\left|N_{t}\right|^{2}\stackrel[n=0]{M-|l-k|}{\sum}\left[C_{n}^{2}C_{n-l+k}^{2}\frac{n!\left(n-l+k\right)!}{\left(n-t-l\right)!}\right]^{1/2},l\leq k.\label{eq:mf}
\end{equation}
Here we have considered $NGBS$ as a particular example of qudit state, whose probability amplitude $C_{n}$ is
defined as

\begin{equation} 
C_{n}=\left[\frac{p}{1+Mq}\frac{M!}{(M-n)!n!}\left(\frac{p+nq}{1+Mq}\right)^{n-1}\left(1-\frac{p+nq}{1+Mq}\right)^{M-n}\right]^{\frac{1}{2}}\label{eq:NGBS}
\end{equation}
Interestingly, for $q=0$ $NGBS$ converted to Binomial state ($BS$), which can further be reduced to Fock state (most nonclassical) and coherent state (most classical) with different limits of depending parameters M and n. Here \(Eq. \) (\ref{eq:NGBS}) is a more general form of probability amplitude of $BS$ so defined as $NGBS$. More details for same are already addressed in our recent paper \cite{mandal2019generalized}. After application
of $'r'$ photon addition and $'t'$ photon subtraction, $NGBS$ is described as
$rNGBS$ and $tNGBS$ respectively. In what follows, we will see that the above analytic expression will essentially
lead to analytic expression for various witness of nonclassicality. In next section, we wish to apply various witness criteria over $rNGBS$ and $tNGBS$ and want to observe the effect of photon addition and subtraction over higher-order nonclassical phenomena.

\section{Nonclassical Properties: Witness Criteria}

A quantum mechanical state $|\psi>$ has $n^{th}$-order nonclassicality
with respect to any arbitrary quantum mechanical operator A, if the
$n^{th}$-order moment of A in that state reduces below to the value of the
$n^{th}$-order moment of A in a poissonian state, i.e. the condition of
$n^{th}$-order nonclassicality with respect to the operator A is
given by $(\text{\textgreek{D}}A)_{|\psi>}^{n}\prec(\text{\textgreek{D}}A)_{|coherent\,state>}^{n}$ where 
$(\Delta A)^{n}$ is the general $n^{th}$-order moment. This is a
general criterion of higher-order nonclassicality in any state originated
by uncertainty principle (see \cite{verma2010generalized}, and references there in). Depending upon the operator form of A, we may
have different criteria of nonclassicality. All these criteria fall
under two categories, 1. Witness 2. Quantifier. Here in section 3,
we wish to study, witness criteria for $rPAQS$ and $tPSQS$. Specifically, we study
HOA, HOS (Hillery type), HOSPS in the next subsections.

\subsection{Higher-order Antibunching}

\begin{figure}
\begin{center}
\begin{tabular}{cc}
    \includegraphics[scale=0.7]{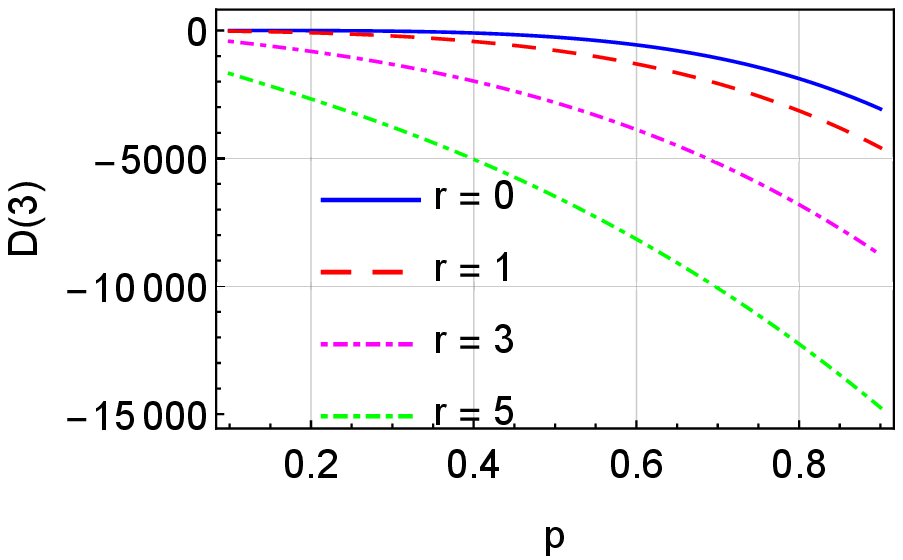} & \includegraphics[scale=0.7]{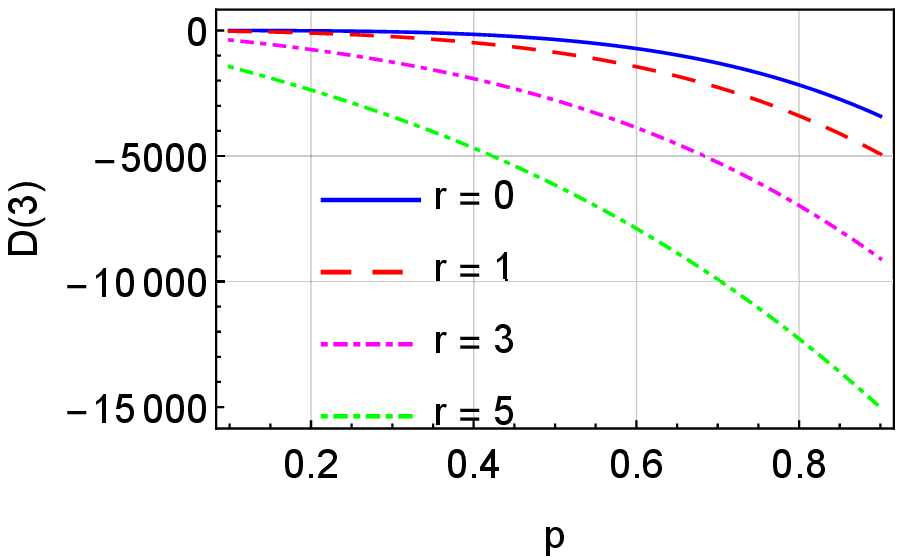}
    \tabularnewline (a) & (b)\tabularnewline
    \end{tabular}

\begin{tabular}{cc}
\includegraphics[scale=0.7]{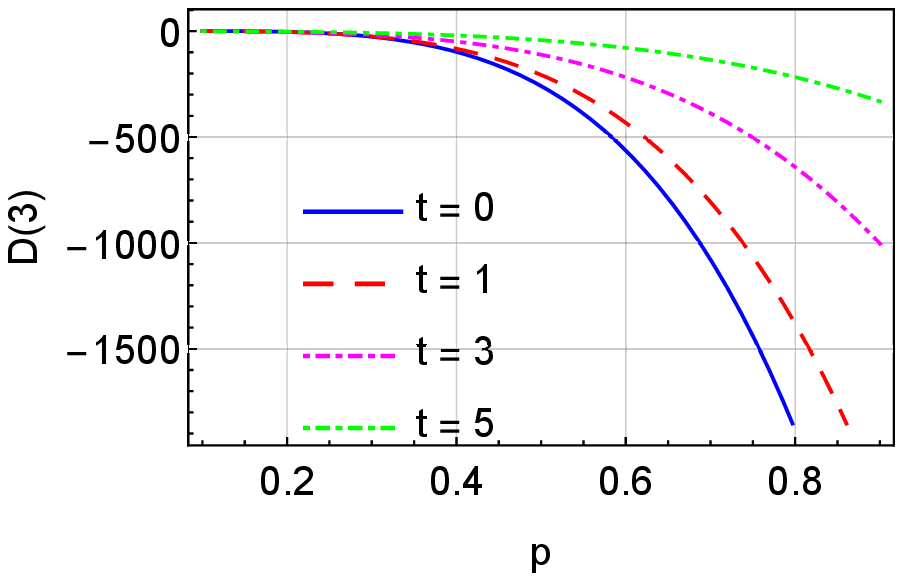} & \includegraphics[scale=0.7]{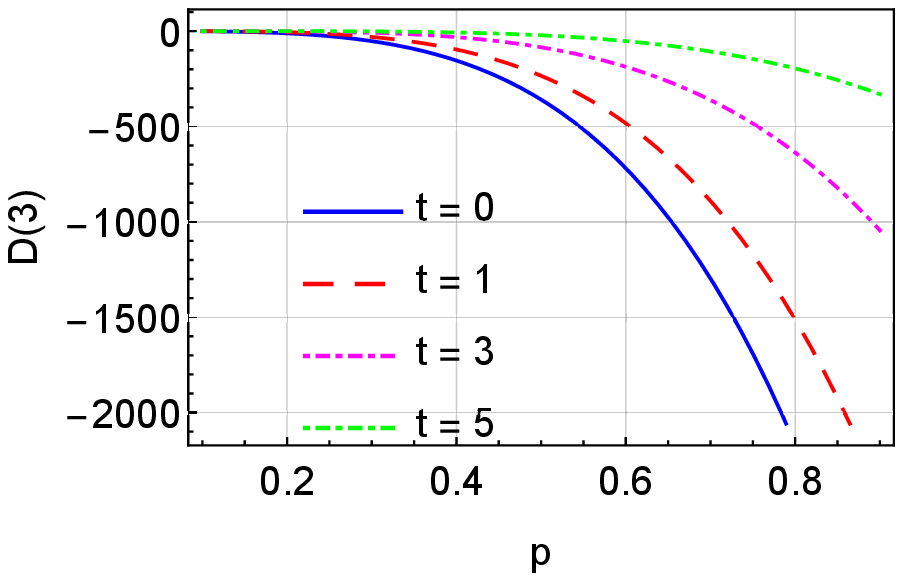}\tabularnewline
(c) & (d)\tabularnewline
\end{tabular}

\caption{\label{fig:Variation-of-HOA1}Variation of HOA for $rNGBS$ and $tNGBS$
is shown here with probability $p$ and for fixed value of order $l=3$
and $M=10$. \(Fig. \) 2(a) and 2(b) are showing HOA for $rNGBS$ with $q=0.01$ and $q=-0.01$ respectively. \(Fig. \) 2(c) and 2(d)
are showing HOA for $tNGBS$ with $q=0.01$ and
$q=-0.01$ respectively. Depth of $HOA$ increases
with photon addition but decreases with photon subtraction in $NGBS$.}
\end{center}
\end{figure}

\begin{figure}
\begin{center}
\begin{tabular}{ccc}
\includegraphics[scale=0.52]{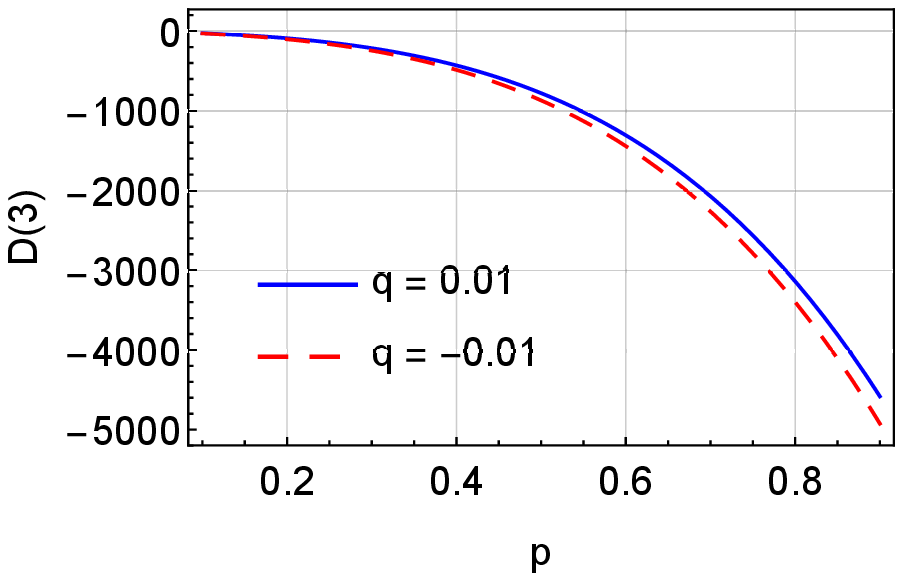} & \includegraphics[scale=0.52]{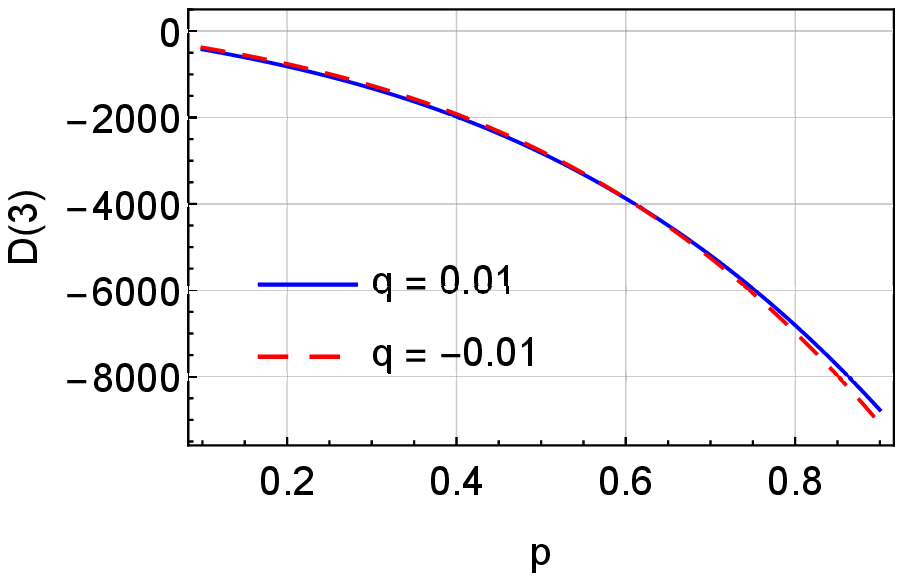} & \includegraphics[scale=0.52]{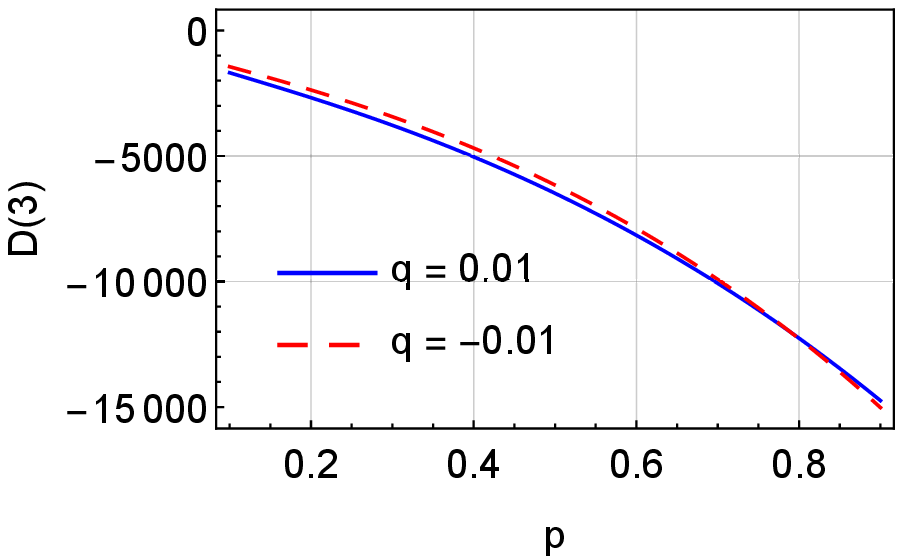}\tabularnewline
(a) & (b) & (c)\tabularnewline
\includegraphics[scale=0.52]{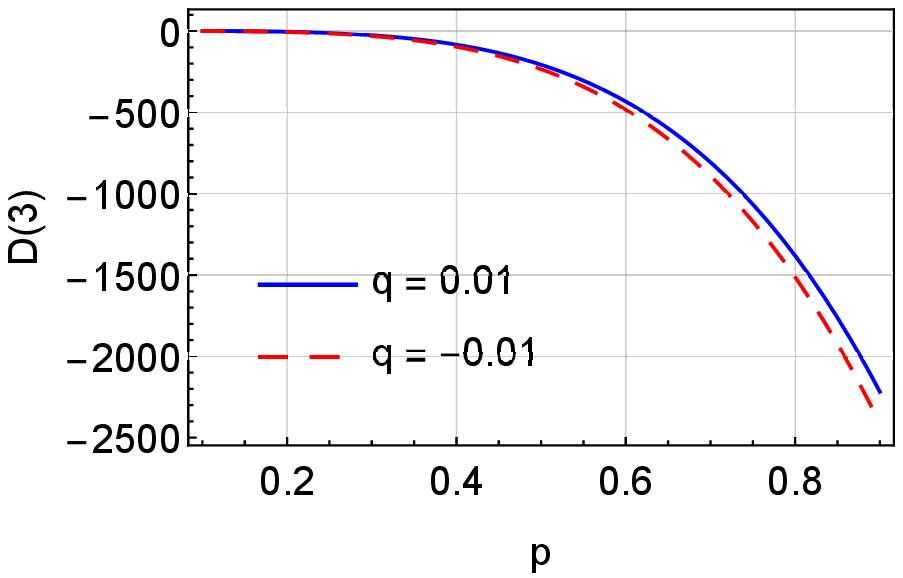} & \includegraphics[scale=0.52]{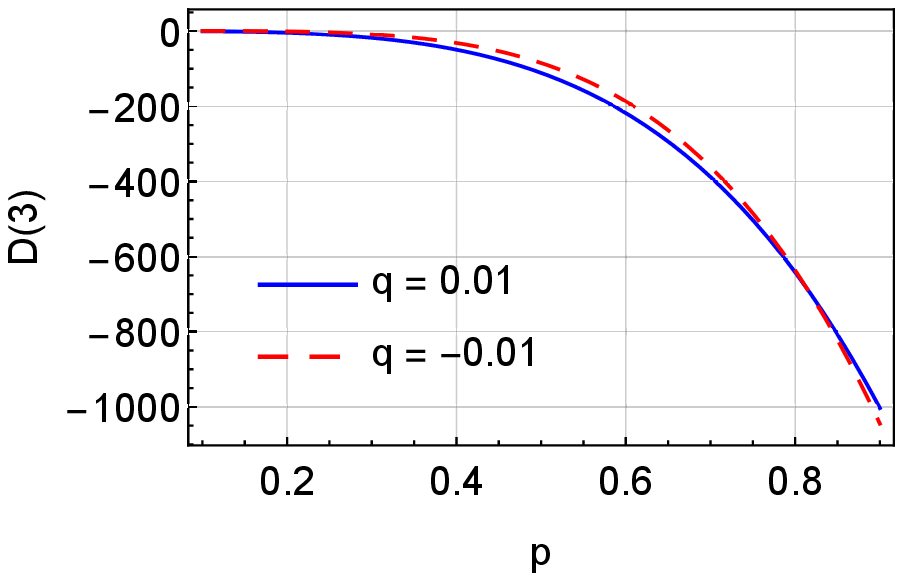} & \includegraphics[scale=0.52]{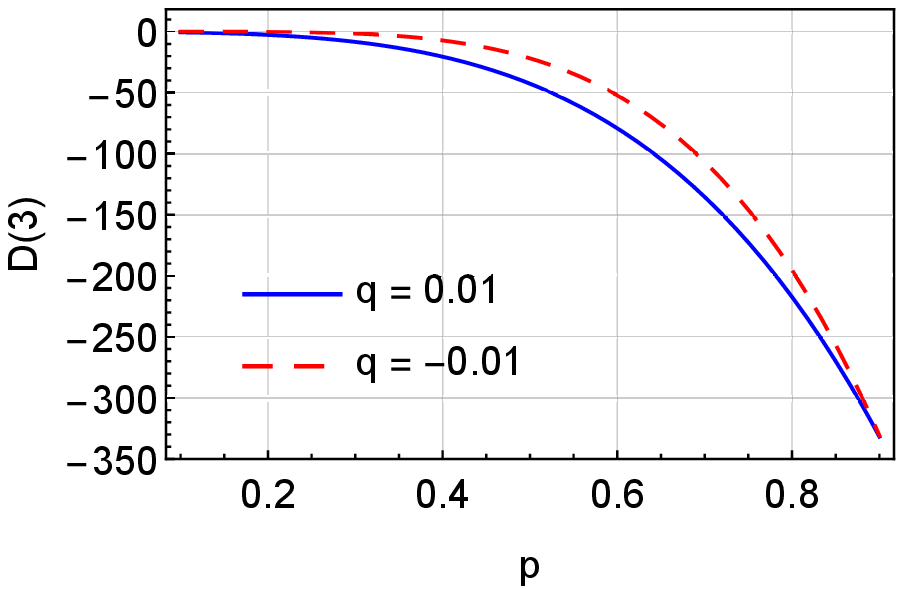}\tabularnewline
(d) & (e) & (f)\tabularnewline
\end{tabular}

\caption{\label{fig:Variation-of-HOA2}(a),(b) and (c) shown HOA for photon added $NGBS$
with $l=3,M=10$ and different values of $q=0.01$ and $q=-0.01,$
$r=1,3,5$ respectively. Similarly (d),(e) and (f) shown HOA for photon subtracted
$NGBS$. It is found that again photon addition is more prominent
than photon subtraction for HOA and state with negative $q$ is more
nonclassical.}
\end{center}
\end{figure}

Phenomenon of antibunching is related to photon statistics
of a state. Using antibunching criterion, one can describe
statistical property of the radiation field. This phenomenon ensures
that in an incident beam of radiation, the probability of getting
two photon simultaneously is less than that of the probability of getting
them separated (one by one). Lee \cite{lee1990higher} in 1990
introduced the criterion of the HOA using the theory
of majorization. However, Lee's criteria was gradually modified by Ba An \cite{an2002multimode} and later by Pathak et al. \cite{pathak2006control}.
From Pathak and Garcia criterion, a quantum state is considered
to be higher-order antibunched state if it satisfies
the following inequality \cite{pathak2006control}

\begin{equation}
D(l)=\langle N^{(l+1)}\rangle-\langle N\rangle^{l+1}=\langle a^{\dagger l+1}a^{l+1}\rangle-\langle a^{\dagger}a\rangle^{l+1}<0,\label{eq:antibunching}
\end{equation}
where $N=a^{\dagger}a$ is the number operator and $N^{(l+1)}=a^{\dagger l+1}a^{l+1}$
is the factorial moment, respectively. For $l=1$, it reduces to the
lower-order antibunching criterion and for $l>1$ it is higher-order
antibunching criterion. Here we have investigated HOA using
the criterion given in \(Eq. \) (\ref{eq:antibunching}). We observed the
existence of HOA in $r$ number of photon added $NGBS$ ($rNGBS$)
and $t$ number of photon subtracted $NGBS$ ($tNGBS$) in \(Fig. \) \ref{fig:Variation-of-HOA1} and \ref{fig:Variation-of-HOA2}.The negative part of the curves ensure
that $rNGBS$ and $tNGBS$ satisfied the inequality (\ref{eq:antibunching}).
So both $rNGBS$ and $tNGBS$ are higher-order antibunched. But in
\(Fig. \) \ref{fig:Variation-of-HOA1}(a) and  \ref{fig:Variation-of-HOA1}(c), we can see that depth
of the HOA witness increases with photon addition and decreases with
photon subtraction. The variation of HOA witness due to $+q$ or $-q$
shown in \(Fig. \)\ref{fig:Variation-of-HOA2} ((a)-(f)).

\subsection{Higher-order Squeezing}

\begin{figure}
\begin{center}
\begin{tabular}{cc}
\includegraphics[scale=0.7]{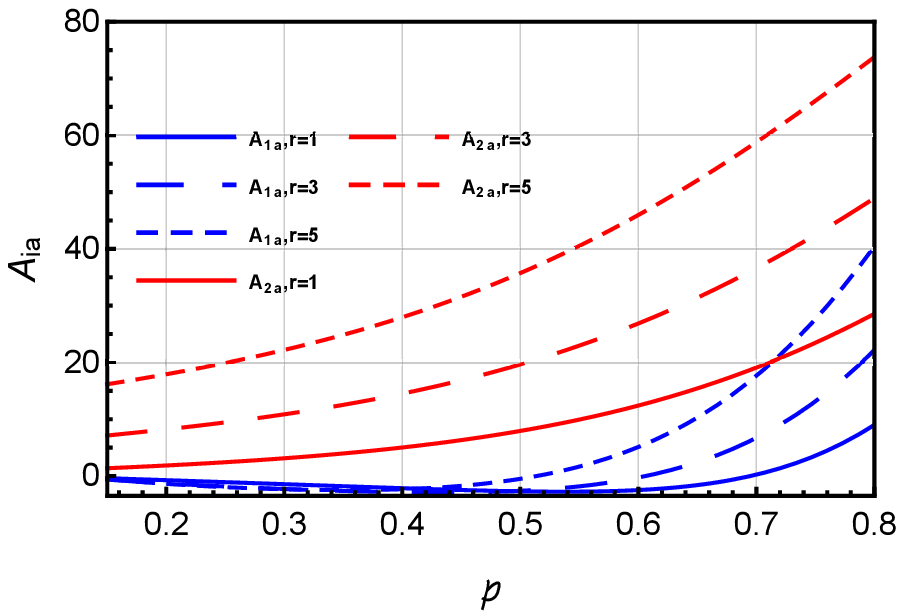} & \includegraphics[scale=0.7]{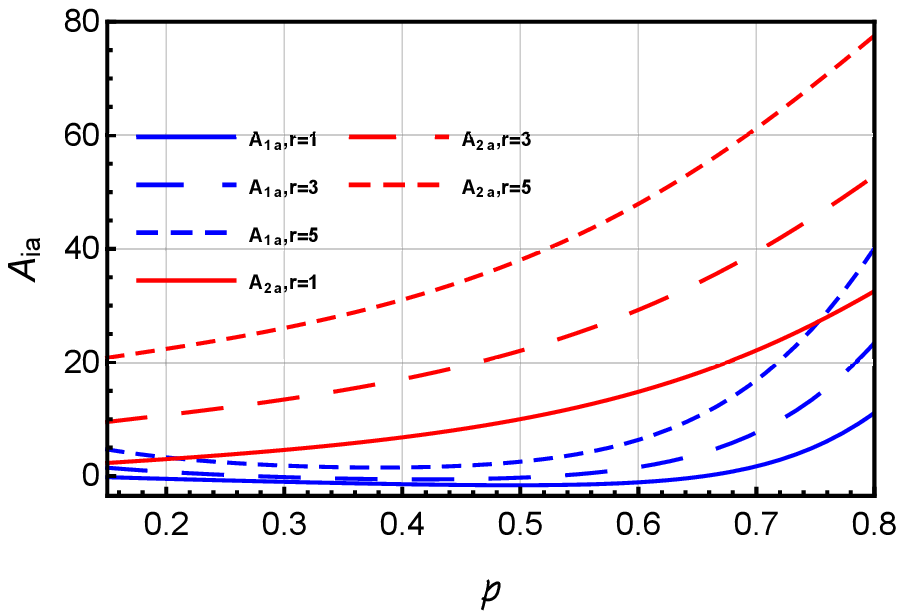}\tabularnewline
(a) & (b)\tabularnewline
\includegraphics[scale=0.7]{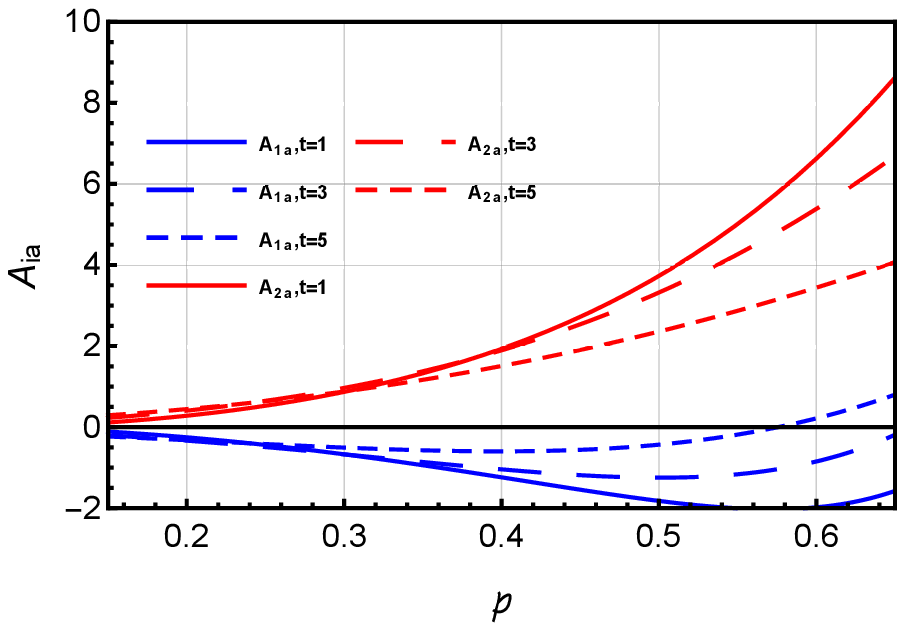} & \includegraphics[scale=0.7]{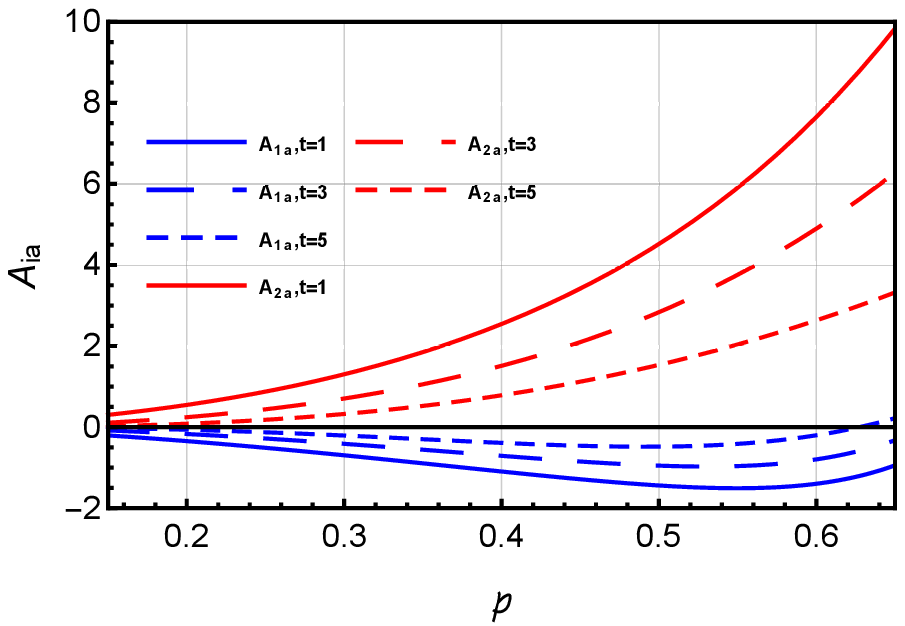}\tabularnewline
(c) & (d)\tabularnewline
\end{tabular}

\caption{\label{fig:Variation-of-HOS1-1}Variation of HOS Hillery type for
photon added and subtracted $NGBS$ is shown here with probability
$p$. 4(a) and 4(b) are showing HOS  for $rNGBS$ with $q=0.01$ and $q=-0.01$ respectively. 4(c) and
4(d) are showing HOS for $tNGBS$ with $q=0.01$
and $q=-0.01$ respectively.}
\end{center}
\end{figure}

The phenomenon squeezing originates from the Heisenberg uncertainty
relation. In which the product of fluctuation of two non commuting
operators in Heisenberg uncertainty relation (uncertainty product)
has a minimum value. At this point both the quadrature variance are
equal. If the variance of one of the quadrature goes below this equal
value (on the cost of increase in other quadrature), the corresponding
quadrature is squeezed. The higher-order counterpart of the squeezing
is higher-order squeezing. In literature we have three types of higher-order
squeezing criteria, Hong-Mandel Squeezing \cite{hong1985generation},
Hillery type squeezing \cite{hillery1987amplitude} and amplitude squared
squeezing in matrix form given by Vogel \cite{shchukin2005nonclassicality}.
Here we have studied Hillery type squeezing which is described as.

\begin{equation}
A_{i,a}=(\Delta Y_{i,a})^{2}-\frac{1}{2}|\langle\left[Y_{1,a},Y_{2,a}\right]\rangle|<0\label{eq:Hill1}
\end{equation}

where $Y_{1,a}=\frac{a^{l}+a^{\dagger l}}{2}$ and $Y_{2,a}=\frac{-i(a^{l}-a^{\dagger l})}{2}$
are amplitude powered quadrature. In this article we have calculated
HOS by amplitude square squeezing i.e. for $l=2$. The result is exhibited
in \(Fig. \) \ref{fig:Variation-of-HOS1-1} (a-d), from where we can easily
conclude that photon subtraction is more effective than photon addition
for getting squeezing and also squeezing is decreasing with number
of photon subtraction.

\subsection{Higher-order sub-Poissonian photon statistics}

\textcolor{black}{Phenomenon of sub-Poissonian photon statistics (SPS) is
again described the statistical property of any qudit state. Lower-order SPS is equivalent to normal antibunching. But the higher-order criterion is different than that of HOA. Higher-order sub-Poissonian photon statistics HOSPS is given by following criterion}

\begin{figure}
\begin{center}
\begin{tabular}{cc}
\includegraphics[scale=0.7]{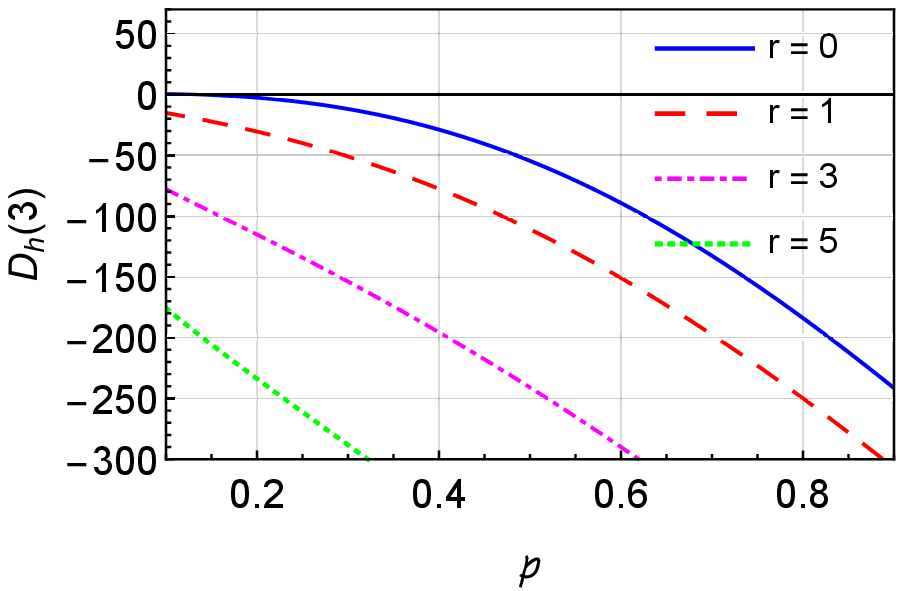} & \includegraphics[scale=0.7]{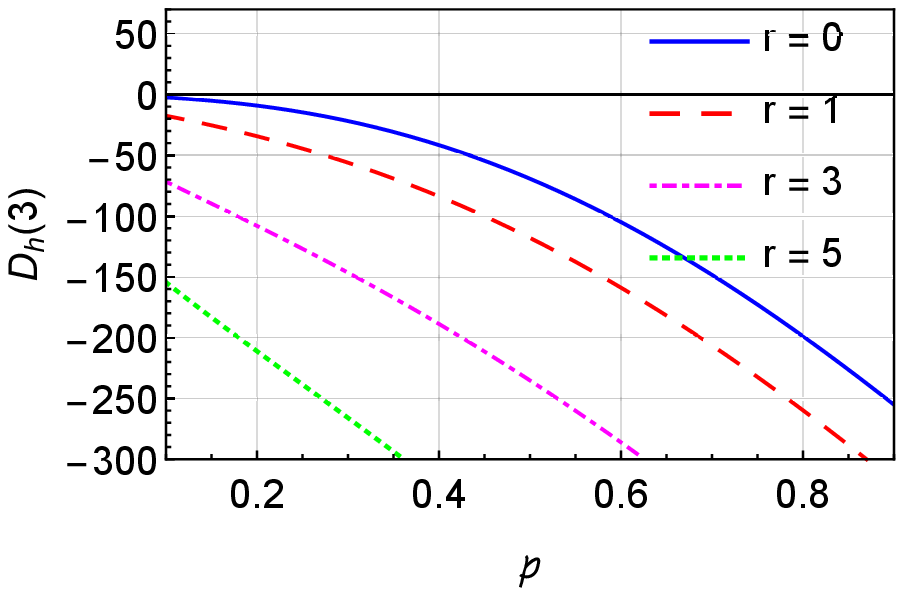}\tabularnewline
(a) & (b)\tabularnewline
\includegraphics[scale=0.7]{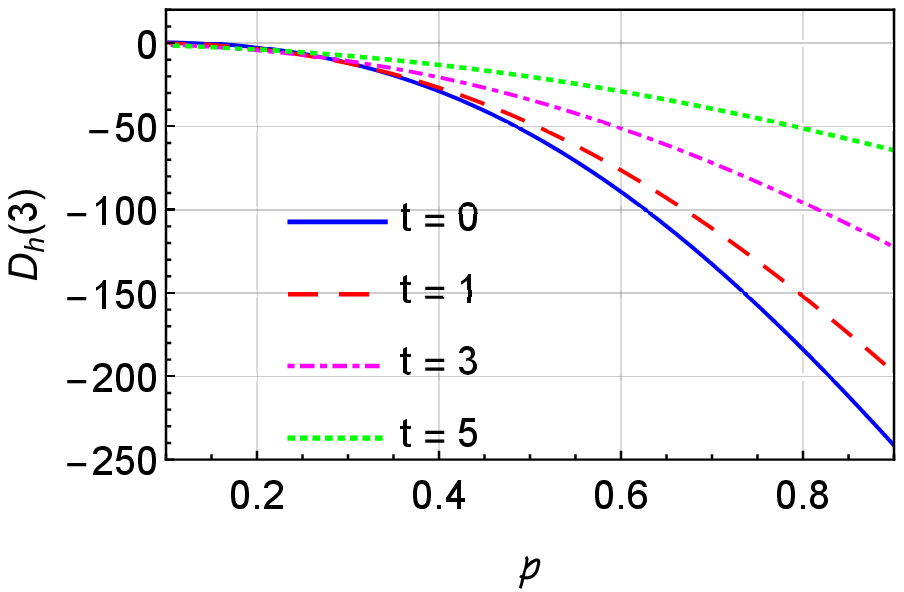} & \includegraphics[scale=0.7]{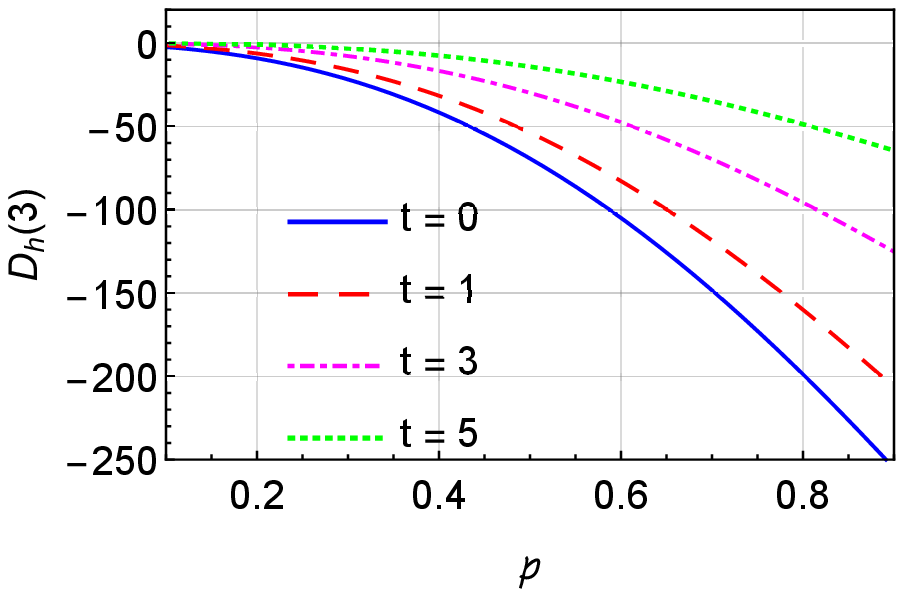}\tabularnewline
(c) & (d)\tabularnewline
\end{tabular}

\caption{\label{fig:Variation-of-Hosps}Variation of HOSPS in $rNGBS$
and $tNGBS$ are shown here with probability $p$ , $M=10$ and
$l=4$. 5(a) and 5(b) are showing existence of HOSPS in $rNGBS$ for $q=0.01$ and  $q=-0.01$ respectively. Similarly 5(C) and 5(d) describe existence of HOSPS in $tNGBS$
with $q=0.01$ and $q=-0.01$ respectively.}
\end{center}
\end{figure}

\textcolor{black}{
\begin{equation}
D_{h}(l-1)=\sum_{r=0}^{l}\sum_{k=0}^{r}S_{2}(r,k)\,^{l}C_{r}(-1)^{r}D(k-1)\langle N\rangle^{l-r}<0\label{eq:HOSPS-1}
\end{equation}
where $S_{2}(r,k)$ is the Stirling number of second kind. The inequality
in \(Eq. \) (\ref{eq:HOSPS-1}) is the condition for the $(l-1)$th order
nonclassicality, and for $l\geq3$ it is the condition for HOSPS. When
higher-order moment of the photon number is less than that of the Poissonian
label i.e., $\langle\left(\Delta N\right)^{l}\rangle<\langle\left(\Delta N\right)^{l}\rangle|_{Poissonian}$}, the state shows
HOSPS. We have obtained an analytic expression
for the inequality in (\ref{eq:HOSPS-1}) by using Eqs. (\ref{eq:mb}),
(\ref{eq:mc}), (\ref{eq:me}) and (\ref{eq:mf}), the corresponding
results are shown in \(Fig.\) \ref{fig:Variation-of-Hosps}((a) to (d))
where the negative parts in figures ensure the HOSPS in $NGBS$. In \(Fig.\) \ref{fig:Variation-of-Hosps}((a)
and (b)), we have observed that the depth of the witness of HOSPS is increasing
with number of photon addition and in \(Fig. \) \ref{fig:Variation-of-Hosps}((c)
and (d)), we have observed that the depth of the witness of HOSPS
decreases with number of photon subtraction.

\subsection{Wigner Function}

\textcolor{black}{In our previous work \cite{mandal2019generalized}
we have derived a compact form of Wigner function of finite dimensional
Fock superposition state (FSS) and also reported the Wigner function
of $NGBS$ which is the special form of $FSS$.}
\begin{equation}
|\psi\rangle=\sum_{n=0}^{N}c_{n}|n\rangle,\label{eq:quantum state1}
\end{equation}
The final expression of the Wigner function of $FSS$ is shown bellow.

\begin{equation}
\begin{array}{lcc}
W(x,p)&=&\frac{1}{\pi^{\frac{1}{2}}}\sum_{n,n^{\prime}=0}^{N}c_{n}^{*}c_{n^{\prime}}b_{n}^{*}b_{n^{\prime}}{\rm e}^{-\left(x^{2}+p^{2}\right)}(-1)^{n^{\prime}}2^{n^{\prime}}n!(ip-x)^{n^{\prime}-n}L_{n}^{n^{\prime}-n}\\
&\times&\left(-2(ip+x)(ip-x)\right)\,\,\,\,\,n\leq n^{\prime}
\end{array}\label{eq:wigner}
\end{equation}

\begin{figure}
	\begin{tabular}{ccc}
		\includegraphics[scale=0.35]{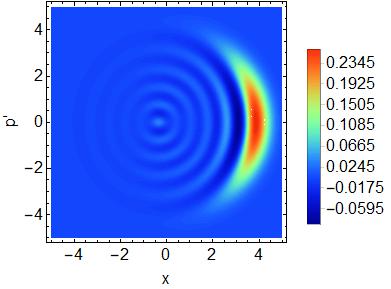} & \includegraphics[scale=0.35]{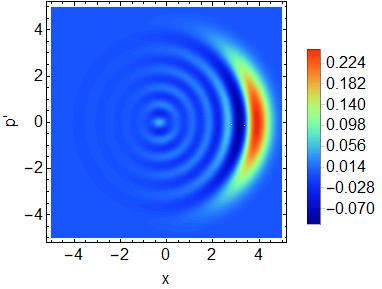} & \includegraphics[scale=0.35]{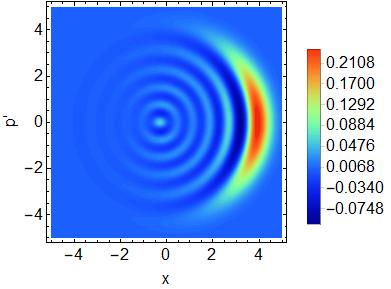}\tabularnewline
		(a) & (b) & (c)\tabularnewline
	\end{tabular}
	
	\caption{\label{fig:Wigner-function-for1} Wigner function for $r$ photon added $NGBS$ with $M=10$ , $q=-0.01$ and $p=0.8$. Different figures (a), (b) and (c) are representing effect of $1$ photon,  $3$ photon and $5$ photon addition respectively.}
\end{figure}

\begin{figure}
	\begin{tabular}{ccc}
		\includegraphics[scale=0.35]{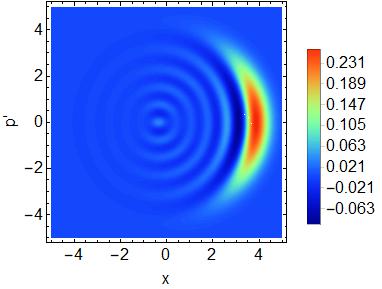} & \includegraphics[scale=0.35]{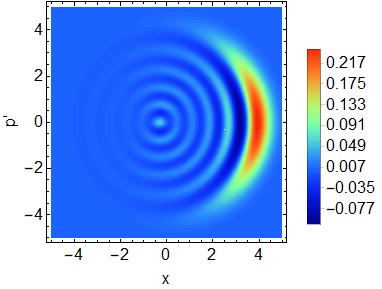} & \includegraphics[scale=0.35]{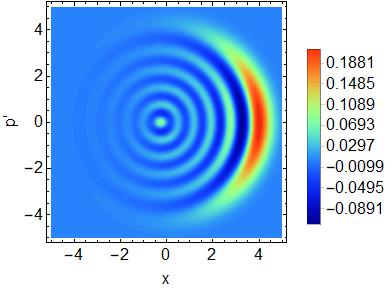}\tabularnewline
		(a) & (b) & (c)\tabularnewline
	\end{tabular}
	
	\caption{\label{fig:Wigner-function-for2} Wigner function for $t$ photon subtracted
		NGBS with $M=10$, $q=-0.01$and $p=0.8$. Different figures (a), (b) and (c) are representing effect of $1$ photon,  $3$ photon and $5$ photon subtraction respectively.}
\end{figure}

\textcolor{black}{By using \(Eq. \)((\ref{eq:mb})(\ref{eq:mc})(\ref{eq:me})
and \(Eq. \)(\ref{eq:mf})), we can easily calculate Wigner function of
$rPAQS$ and $tPSQS$. The obtained Wigner function for $rNGBS$ and $tNGBS$ are shown in \(Fig. \) (\ref{fig:Wigner-function-for1}) and
\(Fig. \)(\ref{fig:Wigner-function-for2}). From results it is clear that
in both the cases, the depth of Wigner function is increasing
with increase in the number of photon addition or subtraction. It is also clear that
photon subtraction is more effective than photon addition in the state. In the next subsection, we wish to show Optical tomograms as it shows probabilistic measurement of Wigner function.}

\subsubsection{Optical Tomogram}

\textcolor{black}{Though there exist some proposals for the direct
measurement of Wigner function as it has probabilistic nature and direct
measurement of Wigner function is not possible. But through data processing,
we may measure it experimentally. Tomogram gives
the probabilistic description of the quantum state which is accessible
for direct measurement. For any quantum state
$|\psi\rangle$, the optical tomogram $w_{|\psi\rangle}\left(X,\theta\right)$
is reported earlier as}

\begin{figure}
\begin{tabular}{ccc}
\includegraphics[scale=0.35]{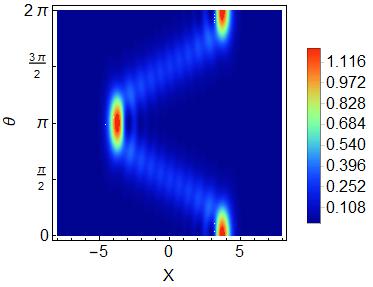} & \includegraphics[scale=0.35]{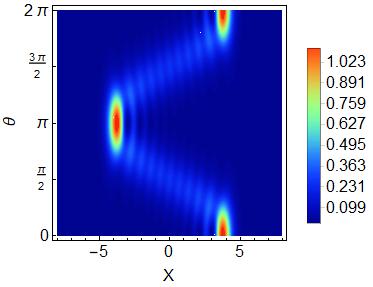} & \includegraphics[scale=0.35]{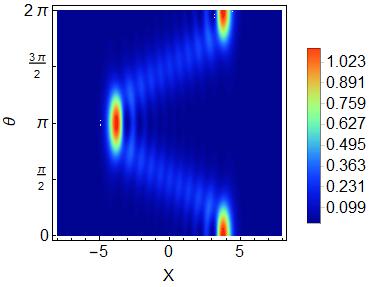}\tabularnewline
(a) & (b) & (c)\tabularnewline
\end{tabular}

\caption{\label{fig:Tomogram-for-photon1}Tomogram for $r$ photon added $NGBS$ with
$M=10,q=-0.01$and $p=0.8$. Different figures (a), (b) and (c) are representing effect of $1$ photon,  $3$ photon and $5$ photon addition respectively.}
\end{figure}

\begin{figure}
\begin{tabular}{ccc}
\includegraphics[scale=0.35]{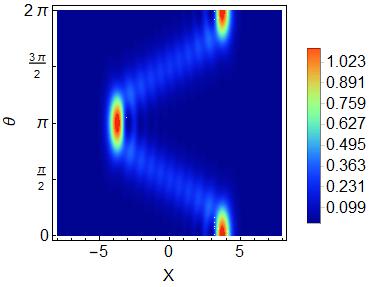} & \includegraphics[scale=0.35]{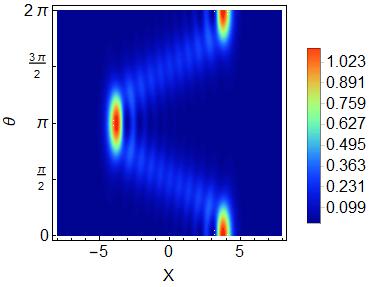} & \includegraphics[scale=0.35]{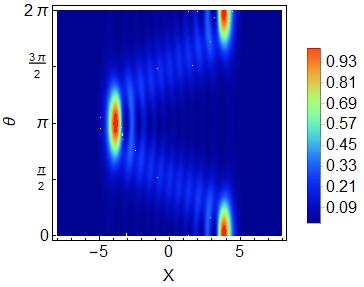}\tabularnewline
(a) & (b) & (c)\tabularnewline
\end{tabular}

\caption{\label{fig:Tomogram-for-photon2}Tomogram for $t$ photon subtracted $NGBS$
with $M=10,q=-0.01$and $p=0.8$. Different figures (a), (b) and (c) are representing effect of $1$ photon,  $3$ photon and $5$ photon subtraction respectively.}

\end{figure}

\textcolor{black}{
\begin{equation}
\begin{array}{lcl}
w_{|\psi\rangle}(X,\theta) & = & \frac{e^{-X^{2}}}{\sqrt{\pi}}\left[\sum_{n=0}^{N}\frac{|c_{n}|^{2}}{2^{n}n!}H_{n}^{2}(X)+\sum_{n<k}\frac{|c_{n}||c_{k}|\cos\left(\left(n-k\right)\theta-\left(\phi_{n}-\phi_{k}\right)\right)}{\sqrt{2^{n+k-2}n!k!}}H_{n}(X)H_{k}(X)\right]\end{array}\label{eq:optical tomogram}
\end{equation}
}\textcolor{black}{where $c_{j}=|c_{j}|e^{i\phi_{j}}$ and $H_{j}$
is the Hermite polynomial of degree $j$. Optical tomogram of $rPAQS$
and $tPSQS$ is calculated by using \(Eq.\) (\ref{eq:mb},\ref{eq:mc},\ref{eq:me},\ref{eq:mf}
and \ref{eq:optical tomogram}). The results are shown in \(Fig. \)(\ref{fig:Tomogram-for-photon1})
and \(Fig. \) (\ref{fig:Tomogram-for-photon2}). In the next section, we wish to present quantifier criterion of nonclassicality as quantitative measurement of the state with the help of Nonclassical Volume.}

\section{Nonclassical Volume: Quantifier criterion}

\textcolor{black}{For quantitative analysis of the state, we study Nonclassical volume which is essentially volume of the negative value region of Wigner function for $r$ photon added $NGBS$ and $t$ photon subtracted $NGBS$. Kenfack and Zyckowski introduced negative volume
of Wigner function as quantifier of nonclassicality in 2004 \cite{kenfack2004negativity}.
In spite of the negative Wigner volume there are several other methods to calculate the
amount of nonclassicality like Hillery's distance-based measure of
nonclassicality \cite{hillery1987nonclassical}, Lee's idea of the nonclassical
depth \cite{lee1991measure}, Asboth et.al idea of using measures
of entanglement as a measure of nonclassicality and Vogel's work \cite{asboth2005computable}.
In this particular measure the volume of the negative part of the
Wigner function is considered as the quantitative measure of nonclassicality. To
be precise, the nonclassical volume associated with a quantum state $|\psi\rangle$
is}

\textcolor{black}{
\[
\delta(\psi)=\int\int\left|W_{\psi}\left(p,q\right)\right|dqdp-1,
\]
}\textcolor{black}{{} where $W_{\psi}\left(p,q\right)$ is the Wigner
function of a quantum state $|\psi\rangle.$ In Table 1,
we have shown variation of $\delta(\psi)$ with number of photon addition
and subtraction. It is clear that nonclassical volume is increased
with the number of photon addition or subtraction. But more effective is photon subtraction as is shown greater volume for the same number of photon. Results are shown in tabular form.}

\begin{center}\begin{tabular}{|>{\centering}p{3cm}|>{\centering}p{3cm}|>{\centering}p{3cm}|}
\hline 
Number of photon added and subtracted & Nonclassical volume for photon added state & Nonclassical volume for photon subtracted state\tabularnewline
\hline 
\hline 
1 & 0.255922 & 0.260153\tabularnewline
\hline 
3 & 0.31384 & 0.353625\tabularnewline
\hline 
5 & 0.363856 & 0.482082\tabularnewline
\hline 
\end{tabular}\end{center}

Table 1: Table is showing Nonclassical volume
of $r$ photon added and $t$ photon subtracted $NGBS$ for $M=10$, $q=-0.01$ and $p=0.8$

\section{Conclusion}
\textcolor{black}{ In the above, we have presented a rigorous 
	study on higher-order nonclassicality of  $rPAQS$ and $tPSQS$ with a general structure which is valid for any qudit states. However, the illustrative examples are given for a particular type of qudit state named as new generalized binomial state $(NGBS)$. First, we describe the
	analytical form of higher-order moment of $rPAQS$ and $tPSQS$ then
	by using general form of moment, we study many criteria of witness
	of nonclassicality like Higher-order Antibunching, Higher-order Squeezing
	(HOS - Hillery type), Higher-order Sub-Poissonian Photon
	statistics (HOSPS), Wigner function and Optical tomogram. Effect of photon addition and subtraction fairly affect nonclassical properties which are shown in the results. In case of HOA
	and HOSPS, depth of nonclassical witness increased by addition of number of photon and decreases by subtraction of photon in NGBS state. In HOS, it is shown
	that photon subtraction is more effective than photon addition. The
	Wigner quasi-probability distribution function of the $rNGBS$ and $tNGBS$
	are reported for photon addition and subtraction. As Wigner function can't be measured directly in general but some can obtained by Optical tomogram with the help of Radon Transform. Here we also computed optical tomogram for $rNGBS$ and $tNGBS$. To make a quantitative analysis, we further calculate amount of nonclassicality in the form of nonclassical volume which is essentially the volume of negative part of the
	Wigner function. It is found that nonclassical volume is increases with the
	number of photon addition or subtraction and photon subtraction gives
	more nonclassicality as a comparison.  From all our observations, it can be concluded
that photon addition and subtraction are useful nonclassicality enhancing or inducing operations. We conclude this paper with the hope that our study will soon be experimentally verified and will also be found to be of use in the study of photon added and subtracted version of other qudit states.}

\section*{Acknowledgement} Authors are thankful to Prof. Anirban Pathak for constant support and encouragement. 
\bibliographystyle{final}
\bibliography{BSartical} 

\end{document}